 \documentclass{emulateapj}

\usepackage{epsfig}

\def\green{f_{_{\rm G}}}

\newcommand{\gapprox}{\lower.4ex\hbox{$\;\buildrel >\over{\scriptstyle\sim}\;$}}
\newcommand{\lapprox}{\lower.4ex\hbox{$\;\buildrel <\over{\scriptstyle\sim}\;$}}
\newcommand{\begeq}{\begin{equation}}
\newcommand{\fineq}{\end{equation}}
\newcommand{\msun}{M_\odot} 
\newcommand{\sig}{\sigma_{_{\rm T}}}

\def\ellprime0{\ell'_0}
\def\colrad{r_0}

\def\Msun{M_\odot}

\def\tauperp{\tau_\perp}

\def\sigpar{\sigma_{_{||}}}
\def\sigperp{\sigma_\perp}

\slugcomment{accepted by ApJ Letters}

\shorttitle{X-Ray Pulsar Spectra}
\shortauthors{Becker \& Wolff}

\begin{document}

\title{SPECTRAL FORMATION IN X-RAY PULSAR ACCRETION COLUMNS}

\author{Peter A. Becker\altaffilmark{1}$^,$\altaffilmark{2}}

\affil{Center for Earth Observing and Space Research,
George Mason University,
Fairfax, VA 22030-4444, USA}

\and

\author{Michael T. Wolff\altaffilmark{3}}
\affil{E. O. Hulburt Center for Space Research,
Naval Research Laboratory,
Washington, DC 20375, USA}

\vfil

\altaffiltext{1}{pbecker@gmu.edu}
\altaffiltext{2}{also Department of Physics and Astronomy,
George Mason University, Fairfax, VA 22030-4444, USA}
\altaffiltext{3}{michael.wolff@nrl.navy.mil}

\begin{abstract}
We present the first self-consistent model for the dynamics and the
radiative transfer occurring in bright X-ray pulsar accretion columns,
with a special focus on the role of the shock in energizing the emerging
X-rays. The pressure inside the accretion column of a luminous X-ray
pulsar is dominated by the photons, and consequently the equations
describing the coupled radiative-dynamical structure must be solved
simultaneously. Spectral formation in these sources is therefore a
complex, nonlinear phenomenon. We obtain the analytical solution for the
Green's function describing the upscattering of monochromatic radiation
injected into the column from the thermal mound located near the base of
the flow. The Green's function is convolved with a Planck distribution
to model the X-ray spectrum resulting from the reprocessing of blackbody
photons produced in the thermal mound. These photons diffuse through the
infalling gas and eventually escape out the walls of the column, forming
the observed X-ray spectrum. We show that the resulting
column-integrated, phase-averaged spectrum has a power-law shape at high
energies and a blackbody shape at low energies, in agreement with the
observational data for many X-ray pulsars.

\end{abstract}


\keywords{pulsars: general --- stars: neutron --- shock waves
--- radiation mechanisms: nonthermal --- methods: analytical
--- X-rays: stars}

\section{INTRODUCTION}

The first pulsating X-ray sources, Her X-1 and Cen X-3, were discovered
in 1971. Since then, over 50 sources have been detected in the Galaxy
and the Magellanic Clouds, with luminosities in the range $L_{\rm X}
\sim 10^{34-38}{\ \rm ergs \ s^{-1}}$ and pulsation periods in the
range $0.1 {\ \rm s} \lapprox P \lapprox 10^3 {\ \rm s}$. In many cases
the X-ray spectra are well fit by a combination of a power-law spectrum
plus a blackbody component (e.g., Coburn et al. 2002; di Salvo et al.
1998; White et al. 1983). Most spectra also display quasi-exponential
cutoffs at $E \sim 20-30\,$keV, and in some cases cyclotron lines are
observed near the cutoff energy, implying magnetic field strengths $\sim
10^{12}$~G. Iron emission lines also appear to be present in certain
sources. The X-ray emission from bright X-ray pulsars is powered by
accretion of material from the ``normal'' binary companion onto the
neutron star.

Efforts to calculate theoretical X-ray pulsar spectra based on either
static or dynamic models have generally yielded results that do not
agree very well with the observed power-law profiles (e.g., M\'esz\'aros
\& Nagel 1985a,b; Nagel 1981; Yahel 1980; Klein et al. 1996), which is
probably due to the neglect of the central role of the shock in
upscattering the radiation. Consequently no satisfactory theoretical
model for the production of X-ray pulsar spectra has yet emerged (Coburn
et al. 2002).

Motivated by the problems with the current models, we reconsider here the
physical picture originally proposed by Davidson (1973), in which the
accreting gas passes through a radiative, radiation-dominated shock
before settling onto the surface of the star. In this situation, most of
the photons are produced in a dense ``thermal mound'' at the base of the
column, just above the stellar surface. These relatively low-energy
photons are upscattered in the shock and eventually diffuse through the
walls of the column, carrying away the kinetic energy of the gas. This
Fermi mechanism characteristically produces a power-law continuum at
high energies.

In this Letter, we demonstrate for the first time how the {\it accretion
shock itself} can directly produce spectra very similar to the continuum
emission seen in many X-ray pulsars. Although this approach has been
suggested before (e.g., Burnard, Arons, \& Klein 1991), it has not been
carried out quantitatively. The work presented here therefore represents
a significant step in the development of a detailed theory for the
spectral formation process in X-ray pulsars.

\section{DYNAMICS OF RADIATION-DOMINATED FLOW}

Radiation pressure plays a crucial role in determining the dynamical
structure of the accretion flows in bright X-ray pulsars. In order for
the accreting gas to be bought to rest at the surface of the star by the
radiation pressure gradient, the X-ray luminosity must satisfy the
constraint $L_{\rm X} \sim L_{\rm crit}$, with the critical luminosity
given by (Becker 1998; Basko \& Sunyaev 1976)
\begin{equation}
L_{\rm crit} \equiv {2.72 \times 10^{37} \sig \over
\sqrt{\sigperp\sigpar}}
\left(M_* \over \msun \right)
\left(r_0 \over R_*\right)
\ {\rm \ ergs \ s}^{-1} \ ,
\label{eq1}
\end{equation}
where $M_*$ is the stellar mass, $R_*$ is the stellar radius, $r_0$ is
the radius of the cylindrical accretion column, $\sig$ is the Thomson
cross section, and $\sigpar$ and $\sigperp$ denote the mean values of
the electron scattering cross sections for photons propagating parallel
or perpendicular to the magnetic field, respectively. The observation of
many X-ray pulsars with $L_{\rm X} \sim 10^{36-38} \, {\rm ergs \,
s}^{-1}$ implies the presence of radiation-dominated shocks close to the
stellar surfaces (White et al. 1983; White et al. 1995).

The accretion flow is assumed to be cylindrical and highly supersonic in
the upstream region. In this case the exact solution for the flow
velocity $v$ is given by (Becker 1998; Basko \& Sunyaev 1976)
\begin{equation}
v(x) = {7 \over 4} \left[1 - \left(7 \over 3\right)
^{-1+x/x_{\rm st}}\right] v_c \ .
\label{eq2}
\end{equation}
The coordinate $x$ increases in the direction of the flow, and $v_c$ is
the flow velocity at the sonic point ($x = 0$), where the Mach number
with respect to the radiation sound speed equals unity. The quantity
$x_{\rm st}$ represents the distance between the sonic point and the
stellar surface, which is related to the other parameters via
\begin{equation}
x_{\rm st} = {r_0 \over 2 \sqrt{3}} \left(\sigperp \over \sigpar
\right)^{1/2} \ln\left(7 \over 3\right) \ .
\label{eq3}
\end{equation}
In the far upstream region, $x \to -\infty$ and $v \to (7/4) \, v_c$
according to equation~(\ref{eq2}). Since the incident (upstream)
velocity is expected to be close to the free-fall velocity onto the
stellar surface, we find that the sonic point velocity $v_c$ is given
by
\begin{equation}
v_c = {4 \over 7} \left(2 \, G M_* \over R_*\right)^{1/2}
\ .
\label{eq4}
\end{equation}

\section{CALCULATION OF THE RADIATION SPECTRUM}

In a steady-state, one-dimensional model, the Green's function
$\green$ representing the response to the injection of $\dot N_0$ photons
per second, each with energy $\epsilon_0$, from a source located at $x=x_0$
satisfies the transport equation
\begin{eqnarray}
v{\partial \green\over\partial x}={dv\over d x}\,{\epsilon\over 3}\,
{\partial \green\over\partial\epsilon} + {\partial\over\partial x}
\left({c\over 3 n_e \sigma_\|}\,{\partial \green\over\partial x}\right)
- {\green \over t_{\rm esc}}
\nonumber
\end{eqnarray}
\begin{equation}
+ {\dot N_0\delta(\epsilon-\epsilon_0)\delta(x-x_0)\over \pi r_0^2
\epsilon_0^2}
- \beta \, v_0 \, \green \delta(x-x_0) \ ,
\label{eq5}
\end{equation}
where $\epsilon$ is the photon energy, $n_e$ is the electron number
density, $v_0 \equiv v(x_0)$ is the flow velocity at the source
location, and $\epsilon^2 \green \, d\epsilon$ gives the number density
of photons in the energy range between $\epsilon$ and $\epsilon +
d\epsilon$. From left to right, the terms in equation~(\ref{eq5})
represent the comoving (advective) time derivative, first-order Fermi
energization (``bulk Comptonization'') in the converging flow, spatial
diffusion parallel to the column axis (i.e., in the $x$-direction),
escape of radiation from the column, photon injection, and the
absorption of radiation at the thermal mound, respectively.

The quantity $t_{\rm esc}$ appearing in equation~(\ref{eq5}) represents
the mean escape timescale for photons to diffuse out through the walls
of the accretion column. If the column cross section is optically thick
to electron scattering as expected, we find that $t_{\rm esc} = r_0 /
w_\perp$, where $w_\perp = c / \tauperp$ is the diffusion velocity
perpendicular to the $x$-axis, and $\tauperp(x) = n_e(x) \, \sigperp \,
\colrad$ is the perpendicular optical thickness of the accretion column
at location $x$. The dimensionless parameter $\beta$ in
equation~(\ref{eq5}) controls the amount of absorption occurring at the
thermal mound. In our application, $\beta$ must be computed
self-consistently since the mound is a blackbody surface that acts
as both a source and a sink of radiation. We will use an energy balance
argument to determine $\beta$ below.

Our primary goal is to calculate the radiation spectrum $f$ resulting from
the injection of blackbody radiation at the thermal mound. Once the
solution for the Green's function $\green$ has been determined, we can
easily compute $f$ by convolving $\green$ with a Planck distribution. If
$\epsilon^2 S(\epsilon) d\epsilon$ denotes the number of photons
injected into the accretion column per second in the energy range
between $\epsilon$ and $\epsilon + d\epsilon$ from the blackbody surface
at $x=x_0$, we have (Rybicki \& Lightman 1979)
\begin{equation}
S(\epsilon) = {2 \, \pi^2 \, r_0^2 \over h^3 \, c^2}
\, {1 \over e^{\epsilon/kT_0} - 1}
\ ,
\label{eq6}
\end{equation}
where $T_0 \equiv T(x_0)$ denotes the gas temperature at the mound
``photosphere.'' The corresponding particular solution for $f$ is then
given by the integral (Becker 2003)
\begin{equation}
f(\epsilon,x) = \int_0^\epsilon {\green(x_0,x,\epsilon_0,\epsilon)
\over \dot N_0} \ \epsilon_0^2 \, S(\epsilon_0) \, d\epsilon_0
\ .
\label{eq7}
\end{equation}
The accretion column in a bright X-ray pulsar is radiation-dominated,
and therefore no net energy can be emitted or absorbed at the thermal
mound. Instead, the absorption of radiation energy is almost perfectly
balanced by the production of fresh blackbody emission. Most of the
energy appearing in the emergent X-rays is transferred to the photons
via collisions with the infalling electrons while the radiation is
diffusing through the column. Based on the energy balance requirement
at the mound ($x=x_0$), we can write the dimensionless absorption constant
$\beta$ in equation~(\ref{eq5}) as
\begin{equation}
\beta = {\int_0^\infty \epsilon^3 \, S(\epsilon) \, d\epsilon
\over \pi r_0^2 \, v_0 U_0}
= {\sigma T_0^4 \over v_0 U_0} \ ,
\label{eq8}
\end{equation}
where $U_0 \equiv U(x_0)$ denotes the radiation energy density at the
top of the thermal mound.

Calculation of the radiation spectrum is simplified considerably if we
work in terms of the new spatial variable $y$, defined by
\begin{equation}
y(x) \equiv
1 - {4 \over 7} \, {v(x) \over v_c}
= \left(7 \over 3\right)^{-1 + x/x_{\rm st}}
\ ,
\label{eq9}
\end{equation}
which has the convenient property that $y \to 0$ in the far upstream
region ($x \to -\infty$), and $y \to 1$ at the surface of the star ($x
\to x_{\rm st}$). Equation~(\ref{eq5}) is separable in energy and space
using the functions
\begin{equation}
f_\lambda(\epsilon,y) = \epsilon^{-\lambda} \, g(\lambda,y)
\ ,
\label{eq10}
\end{equation}
where $\lambda$ is the separation constant and the spatial function $g$
satisfies the differential equation
\begin{eqnarray}
y(1-y)\,{d^2 g \over dy^2} + {1 - 5 \, y \over 4}\,{d g \over dy}
+ \left[{\lambda \over 4} + {y - 1 \over 4 \, y} \right]
\, g
\nonumber
\end{eqnarray}
\begin{equation}
\phantom{SPACE}
= {3 \, \beta \, v_0 \, \delta(y-y_0) \over 7 \, v_c} \, g \ ,
\label{eq11}
\end{equation}
with solutions given by
\begin{equation}
g(\lambda,y) = \cases{
\varphi_1(\lambda,y) \ , & $y \le y_0$ \ , \cr
B \, \varphi_2(\lambda,y) \ , & $y \ge y_0$ \ , \cr
}
\label{eq12}
\end{equation}
where $y_0 \equiv y(x_0)$ and $B = \varphi_1(\lambda,y_0) /
\varphi_2(\lambda,y_0)$ to ensure continuity of $g$ at $y=y_0$ as
required. The functions $\varphi_1$ and $\varphi_2$ are combinations of
hypergeometric functions satisfying the appropriate upstream and
downstream boundary conditions (Becker \& Wolff 2005). The system must
also satisfy the eigenvalue relation
\begin{equation}
\varphi_1 \, {\partial \varphi_2 \over \partial y}
- \varphi_2 \, {\partial \varphi_1 \over \partial y}
- {3 \, \beta \, v_0 \, \varphi_1 \, \varphi_2 \over 7 \,
v_c \, y \, (1-y)} \ \Bigg|_{y=y_0} = 0 \ .
\label{eq13}
\end{equation}
Equation~(\ref{eq13}) determines the eigenvalues $\lambda_n$ and the
associated set of orthogonal eigenfunctions $g_n(y) \equiv
g(\lambda_n,y)$. The first eigenvalue $\lambda_0$ is especially
important because it characterizes the shape of the high-energy spectrum
emerging from the column.

\begin{figure}
\begin{center}
\epsfig{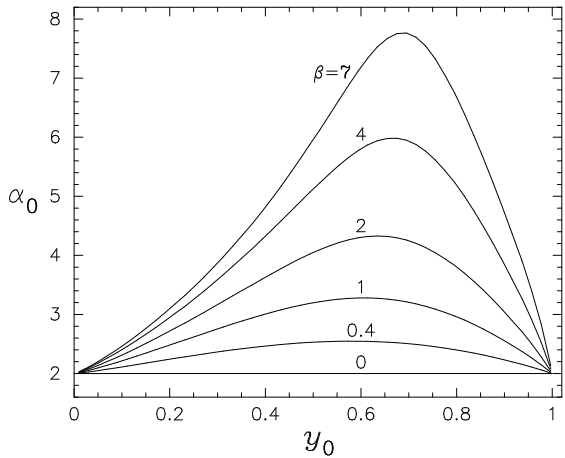}
\end{center}
\caption{High-energy power-law spectral index $\alpha_0$ of the emitted
photon number distribution $\Phi_\epsilon$ (eq.~[\ref{eq15}]) plotted
as a function of the source location $y_0$ for the indicated values of
the absorption parameter $\beta $. Note the steepening of the radiation
spectrum that occurs when $\beta$ is increased for a fixed value of
$y_0$, which reflects the decreasing residence time for the photons in
the plasma due to the enhanced absorption.}
\end{figure}

Once the eigenvalues and eigenfunctions are determined, the solution for
the Green's function can be expressed as the infinite series
\begin{equation}
\green(y_0,y,\epsilon_0,\epsilon) = \sum_{n=0}^\infty \, C_n \,
\left(\epsilon \over\epsilon_0\right)^{-\lambda_n} g_n(y) \ ,
\label{eq14}
\end{equation}
where the expansion coefficients $C_n$ depend on the strength and
location of the source. The phase-averaged spectrum emitted by the X-ray
pulsar is computed by integrating the particular solution $f$ over the
height of the column to obtain
\begin{equation}
\Phi_\epsilon(\epsilon)
= \pi r_0^2 \, \epsilon^2 \int_{-\infty}^{x_{\rm st}}
t_{\rm esc}^{-1} \, f(\epsilon,x) \ dx \ ,
\label{eq15}
\end{equation}
where $\Phi_\epsilon(\epsilon)\,d\epsilon$ gives the number of photons
escaping from the column per unit time with energy between $\epsilon$
and $\epsilon+d\epsilon$, and $f$ is computed using
equation~(\ref{eq7}). The associated high-energy photon spectral index,
$\alpha_0 \equiv \lambda_0-2$, is plotted in Figure~1 as a function of
the model parameters $\beta$ and $y_0$. Note that a broad range of
indices can be produced, although we must have $\alpha_0 > 2$ in order
to avoid an infinite photon energy density since the model considered
here does not include a high-energy cutoff.

\section{ASTROPHYSICAL APPLICATIONS}

For given values of the stellar mass $M_*$ and the stellar radius $R_*$,
our model has three free parameters, namely the column radius $r_0$, the
temperature at the top of the thermal mound $T_0$, and the accretion
rate $\dot M$. The first step in calculating the photon number spectrum
$\Phi_\epsilon$ is to determine the value of $x_0$ (or, equivalently,
$y_0$) in terms of $r_0$, $T_0$, and $\dot M$, which fixes the location
of the thermal mound. The velocity and density at the top of the mound,
$v_0$ and $\rho_0$, respectively, are related via the continuity equation
$\dot M = \pi r_0^2 \, v_0 \, \rho_0$. The density $\rho_0$ can be
calculated by setting the Rosseland mean of the free-free absorption
optical thickness across the column equal to one, which yields for
pure, fully-ionized hydrogen (Rybicki \& Lightman 1979)
\begin{equation}
\rho_0
= J \, v_0^{-1}
= 4.05 \times 10^{-12} \ T_0^{7/4} \ r_0^{-1/2}
\ ,
\label{eq16}
\end{equation}
in cgs units, where we have set the Gaunt factor equal to unity and
introduced the conserved mass flux $J \equiv \dot M/(\pi r_0^2)$. Once
$v_0$ is computed, we can combine equations~(\ref{eq4}) and (\ref{eq9})
to obtain
\begin{equation}
y_0
= 1 - v_0 \left(2 \, G M_* \over R_* \right)^{-1/2} \ ,
\label{eq17}
\end{equation}
which determines the location of the mound. For typical X-ray pulsar
parameters, we find that $y_0 \lapprox 1$, and therefore the mound is
situated very close to the surface of the star as expected. The final
parameter that needs to be evaluated is $\beta$. Using
equations~(\ref{eq4}) and (\ref{eq9}), along with equation~(3.11) from
Becker (1998), we find that the spatial variation of the radiation
energy density is given by
\begin{equation}
U(y) = {3 \dot M y \over \pi r_0^2}
\, \left(2 \, G M_* \over R_*\right)^{1/2} \ .
\label{eq18}
\end{equation}
Combining equations~(\ref{eq8}) and (\ref{eq18}), we find that
\begin{equation}
\beta
= {\pi r_0^2 \, \sigma T_0^4 \over 3 \dot M \, y_0 \, v_0}
\left(2 \, G M_* \over R_* \right)^{-1/2} \ .
\label{eq19}
\end{equation}
Taken together, equations~(\ref{eq4}), (\ref{eq16}), (\ref{eq17}), and
(\ref{eq19}) allow the determination of the four model parameters $v_c$,
$v_0$, $y_0$, and $\beta$ in terms of $R_*$, $M_*$, $r_0$, $T_0$, and
$\dot M$. This closes the system and facilitates the calculation of the
emergent photon number spectrum $\Phi_\epsilon$ using
equation~(\ref{eq15}).

\begin{figure}
\begin{center}
\epsfig{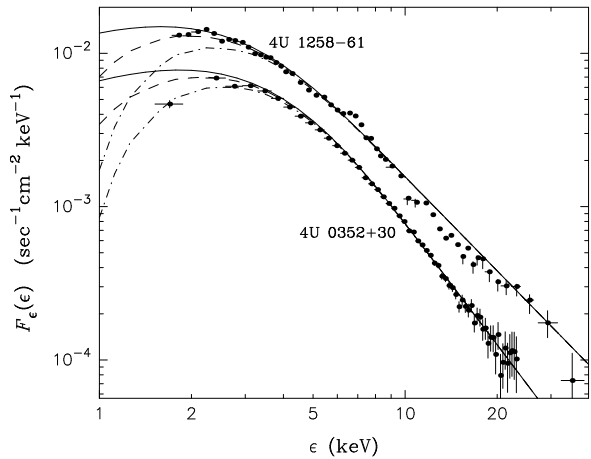}
\end{center}
\caption{Theoretical count-rate flux distribution $F_\epsilon(\epsilon)
\equiv \Phi_\epsilon/(4 \pi D^2)$ computed using various amounts of
interstellar absorption are compared with the observed X-ray spectra of
4U~1258-61 and 4U~0352+30. The column densities used are $N_{\rm H} = 0$
({\it solid lines}); $N_{\rm H} = 3 \times 10^{21}\,{\rm cm} ^{-2}$
({\it dashed lines}); and $N_{\rm H} = 9 \times 10^{21}\,{\rm cm}^{-2}$
({\it dot-dashed lines}). The other parameters for the theoretical
models are given in the text.}
\end{figure}

In Figure~2 we plot the theoretical predictions for the observed
count-rate flux distribution, $F_\epsilon \equiv \Phi_\epsilon/(4 \pi
D^2)$, along with the observed spectra for the X-ray pulsars 4U~1258-61
(GX~304-1) and 4U~0352+30 (X~Persei), respectively, where
$\Phi_\epsilon$ is computed using equation~(\ref{eq15}) and $D$ is the
distance to the source. The 4U~1258-61 data was reported by White et al.
(1983), and the 4U~0352+30 data is the result of XSPEC analysis of an
archival RXTE observation taken in July 1998 and reported by
Delgado-Mart\'i et al. (2001). The observational data represent the
de-convolved (incident) spectra, and consequently there is some weak
model dependence in these results. In both cases we set $M_* = 1.4 \,
\Msun$ and $R_* = 10\,$km. For 4U~1258-61, the model parameters used are
$r_0=6\,$km, $T_0=7.3 \times 10^6\,K$, $\dot M=2.69 \times 10^{16}\,{\rm
g \, s}^{-1}$, and $D=2.5\,$kpc, which yields for the photon index
$\alpha_0 = 2.04$. For 4U~0352+30, we set $r_0=1.3\,$km, $T_0=9 \times
10^6\,$K, $\dot M=3.23 \times 10^{13}\,{\rm g \, s}^{-1}$, and $D =
0.35\,$kpc, which yields $\alpha_0 = 2.64$. The distance estimates are
taken from Negueruela (1998), and various amounts of interstellar
absorption are included as indicated in the plots. The second source
(X~Per) is included due to its low luminosity, which presents an
interesting test for the model. Although the results presented here are
not fits to the data, we note that the general shape of the spectra
predicted by the theory agrees fairly well with the observations for
both sources, including the turnover at low energies and the power law
at higher energies. Several other sources yield similar agreement. In
our model, the turnover at $\sim 2\,$keV is due to Planckian photons
that escape from the accretion column without experiencing many
scatterings. This effect will tend to reduce the amount of absorption
required to fit the observational data.

\section{CONCLUSIONS}

We have shown that a simplified model comprising a radiation-dominated
accretion column with a blackbody source/sink at its base and a
radiative shock is able to reproduce the power-law spectra observed in
accretion-powered X-ray pulsars with a range of luminosities and
power-law indices. The power-law plus blackbody form for pulsar spectra
has previously been adopted in a purely ad hoc manner, but the new model
described here provides a firm theoretical foundation for this general
result. Our work represents the first ab initio calculation
of the X-ray spectrum associated with the physical accretion scenario
first suggested by Davidson (1973).

Our goal in this Letter is to highlight the direct role of the accretion
shock in producing the observed spectra, and therefore we have not
included many other details that are certainly relevant in some sources.
For example, the observed spectrum of Her~X-1 has a very flat power-law
shape up to a sharp exponential cutoff. The photon index in this case is
$\alpha_0 \sim 1$, which is outside the range allowed by our model since
no high-energy cutoff is included here. In order to explain this shape,
additional effects need to be included such as thermal Comptonization,
cyclotron features, iron emission lines, and the possibility of an
ultrasoft component produced by the accretion disk or the surrounding
stellar surface. The inclusion of cyclotron cooling is particularly
interesting since it connects the magnetic field with the flow dynamics
by altering the pressure distribution, which can lead to further changes
in the spectrum. We intend to explore these issues in future work.

The authors are grateful to the anonymous referee for several
suggestions that led to improvements in the manuscript. PAB would also
like to acknowledge the generous support from the Office of Naval
Research.

\end{document}